\documentclass[fleqn,usenatbib]{mnras}

\usepackage[T1]{fontenc}

\DeclareRobustCommand{\VAN}[3]{#2}
\let\VANthebibliography\thebibliography
\def\thebibliography{\DeclareRobustCommand{\VAN}[3]{##3}\VANthebibliography}

\usepackage{soul}

\usepackage{graphicx}	
\usepackage{amsmath}	
\usepackage{amssymb}	

\newcommand{\mph}{$h^{-1}\rm Mpc\,$}
\newcommand{\msol}{$\rm M_{\odot}\,$}

\title[Environmental effects on void galaxies]{Local and large-scale effects on the astrophysics of \textbf{void galaxies}}

\author[Rodr\'iguez M et al.]{
Agust\'in M. Rodr\'iguez-Medrano$^{1,2}$\thanks{E-mail: arodriguez@unc.edu.ar},
Dante J. Paz$^{1,2}$,
Federico A. Stasyszyn$^{1,2}$, 
Facundo Rodr\'iguez$^{1,2}$,
\newauthor
Andr\'es N. Ruiz$^{1,2}$ and
Manuel Merch\'an$^{1,2}$
\\
$^{1}$Instituto de Astronomía Teórica y Experimental, UNC-CONICET, Laprida 854, X5000BGR Córdoba, Argentina\\
$^{2}$ Universidad Nacional de Córdoba. Observatorio Astronómico de Córdoba. Laprida 854, X5000BGR, Córdoba, Argentina\\
}

\date{Accepted XXX. Received YYY; in original form ZZZ}

\pubyear{2023}

\begin{document}
\label{firstpage}
\pagerange{\pageref{firstpage}--\pageref{lastpage}}
\maketitle

\begin{abstract}
Galaxies in cosmic voids have been reported with properties related to a delayed evolution compared to the rest of the Universe.
These characteristics reflect the interaction of galaxies with the environment. 
However, it is not clear the degree of influence of the large-scale structure on the properties of void galaxies or, if these are only influenced by the low local density around them typical of these regions.
In this article we identified cosmic voids in the SDSS-DR16 and studied the g-r colour, star formation rate, and concentration of galaxies.
We identified galaxy groups to characterise the local environment and studied the properties of galaxies as a function of total and stellar masses, separately analyzing those in voids and the general sample.
Our results show that galaxies that inhabit haloes of a given mass (below $\sim \,10^{13.5}M_{\odot}$), are bluer, have a higher star formation rate and are less concentrated when the host halo is inside voids compared to other regions. 
For larger halo masses, the trend disappears.
We also analyse whether the properties of galaxies are sensitive to the type of voids that inhabit. 
This is done by separating voids embedded in overdense regions (S-type) from those that asymptotically converge to the average density of the universe (R-type).
We found that galaxies in R-type voids are bluer, with higher SFR and less concentration than in S-type voids.
Our results indicate some degree of correlation of galaxy properties with the large-scale environment provided by voids, suggesting possible second-order mechanisms in galaxy evolution.

\end{abstract}




\section{Introduction}

Galaxy observations collected in large-scale structure surveys and the advent of numerical simulations in recent decades have provided invaluable clues about the processes of galaxy formation and evolution.
The dependence of galaxy properties on the environment reveals to a certain extent the processes that have occurred in the history of galaxies and how these are linked to the large-scale structure of the Universe \citep{Dressler1980, Postman1984}.
The study of this dependence and the mechanisms that are triggered in the evolution of galaxies is and has been of great interest in modern astronomy
\citep[see for instance][and references there in]{Coldwell2002,Martinez2006,Omill2008,Alpaslan2015,Duplancic2018,Duplancic2020}

Structures such as nodes and filaments may be environments in which galaxies are strongly affected by tidal effects
\citep[e.g. ][and references there in]{Rost2020}.
On the other hand, the void regions may be sites where galaxies evolve slowly \citep{Rojas2005, vonBenda2008, Martizzi2020}, mainly due to local effects and being more representative of early-stage galaxies.
The differences in the astrophysical properties of galaxies inside or outside voids have been studied in several articles.
Photometric and spectroscopic studies reported that void galaxies are blue and star forming
\citep{Grogin1999,Rojas2004, Patiri2006b, vonBenda2008, Wegner2008,Tavasoli2015, Moorman2016, Ceccarelli2022, Jian2022}
while also, it was found that stellar populations in void galaxies are younger than in the general Universe \citep{Rojas2005}. 
These results support the idea of a slower evolution of structures inside voids in regard to the field galaxies.
Although all these results suggest a different evolutionary path for galaxies in voids, it is not clear if it is only a trend due to the low-density local environment that surrounds the galaxies, or if it reflects the influence of cosmic voids (understood as large-scale environment) on the properties of galaxies.
In the case of morphological aspects, the properties of the galaxies also show signs of slower evolution, being reported as late-type galaxies \citep{Huchtmeier1997,Hoyle2005,Ricciardelli2017}.

As shown in numerous studies, the properties of galaxies are strongly affected by the local density \citep[see for instance][]{Hashimoto1998,Lewis2002,Blanton2005,Cooper2005,Padilla2010}. 
Nevertheless, \cite{Kauffman2004} found that large-scale density (r > 1Mpc) has no impact on the star formation history of galaxies with a fixed local density. Conversely, studies such as \citet{Balogh2004a} found that large-scale density does impact the star formation history of galaxies. This highlights the ambiguity surrounding the effect of large-scale density on the properties of galaxies.
On the other hand, \citet{Ceccarelli2008,Ceccarelli2012} studied void galaxies in equal-density environments and found evidence that the large-scale void environment affects galaxies by increasing the relative fraction of blue and actively star-forming galaxies relative to the field.

The way in which galaxies populate the haloes also shows dependence on the environment. This effect, quantified in terms of the halo occupation distribution (HOD), was studied in numerical simulations \citep{Alfaro2020} and observations \citep{Alfaro2022} where the authors found that haloes of the same mass in voids contain fewer galaxies than haloes in denser environments. 
One aspect that remains unclear is the variation of the stellar-halo mass relation of central galaxies with the environment. 
In a number of studies with numerical simulation it has been found that haloes in voids have less massive central galaxies compared with the mean of the Universe \citep{Alfaro2020,Habouzit2020,Rosas-Guevara2022}. However, observational studies such as \citet{Douglass2019} have failed to detect signs of a different stellar-to-halo relation for void environments. 
Therefore, the lack of clarity on these issues motivates us to study these relationships further in this work.

Moreover the inner voids regions, the surrounding of cosmic voids also appear to influence the properties of their galaxies \citep{Thompson2022,RodriguezMedrano2022}. Voids with smaller radii appear to have an excess of red and elliptical galaxies in their walls regarding the biggest voids \citep{Ricciardelli2014,Ricciardelli2017}. This results can be associated with the dynamical difference  found in \citet{Ceccarelli2013} where it was reported that typically the smallest voids have contracting walls and the biggest expanding voids.

The present work aims to study if cosmic void structures affect the properties of the galaxies or if void galaxies only respond to local density effects. We segregate the galaxies according to their local density and their location regarding cosmic voids and study their colour, star formation and concentration. We choose these properties because they are related to the evolution stage of the galaxies where the void environment can be an influence \citep{Alfaro2020,RodriguezMedrano2022}. 

This paper is organised as follows: in section 
\ref{sec_data} we present the data used in this work and the algorithm used to identify voids and galaxy groups. In section \ref{sec_voidgalaxies} we show the results found for the galaxies that inhabit the inner voids regions. In section \ref{sec_nearvoids} we extend the analysis to galaxies in the near void regions. Finally, in section \ref{sec_discusion} we discuss our results and in section \ref{sec_conclusions} we present our main conclusions.

\section{Data}
\label{sec_data}
 
With the aim to differentiate between local and large scale effects on galaxies we identified groups and voids in a volume limited sample of galaxies in the Sloan Digital Sky Survey Data Release 16 \citep[SDSS DR16,][]{DR16}. 
We selected galaxies with apparent magnitude in the r-band $\rm m_r<17.77$ and redshift in the range of $0.02<z<0.1$.
Throughout this paper we have used stellar masses and star formation rates measurements from the MPA-JHU catalogue \citep{Kauffman2003}. 

\subsection{Void-identification}
\label{sec_void}
We identified voids in a volume limited sample at $z=0.1$ with absolute magnitudes in the r-band brighter
than $M_{\rm r} - 5\log_{10}(h) < -20$ using the algorithm described in \citet{Ruiz2015}. 
In a nutshell, the voids are defined as spherical regions, without overlapping
and with an integrated density contrast ($\Delta$) bellow a threshold of $-0.9$. 

The application on observational data is described in detail in \citet{Ruiz2019}, but here we briefly describe the method.  
In a first step the algorithm perform a Voronoi tesellation on the tracer galaxies.
The density of each cell is defined as $\rm \rho_{cell}=1/V_{cell}$ where $\rm V_{cell}$
is its volume. The density contrast in each cell is computed as:
\begin{equation}
\rm \delta_{cell} = \frac{\rho_{cell}}{\hat{\rho}}-1   
\end{equation}
where $\hat{\rho}$ is the mean density of tracers.
This density is obtained by dividing the total number of galaxies in the sample by
its comoving volume. The volume is estimated using the comoving
distance at the redshift threshold, $z=0.1$, and the solid angle of the survey.
This last value was calculated using an angular mask that considers the gaps
and the limits of the survey \citep[see details in][]{Ruiz2019}. 

\begin{figure}
	\includegraphics[width=\columnwidth]{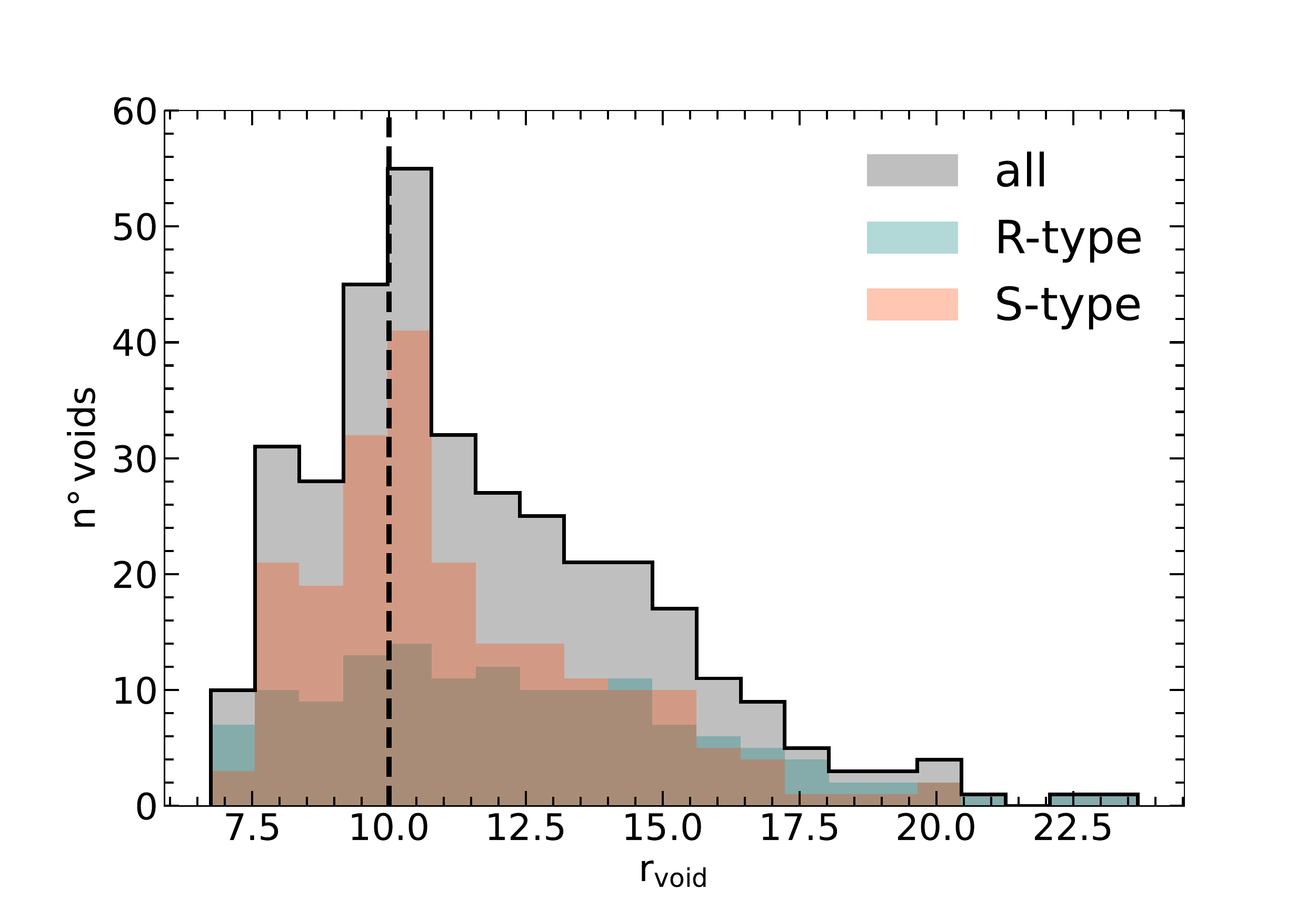}
    \caption{ The grey shaded area corresponds to the histogram of radii of the 
    void sample used in this work. The vertical dashed black line indicate
    the shot noise radius ($10\,h^{-1}\rm Mpc$).
    The coloured shaded areas correspond to the distribution of S-type and
    R-type voids (see section \ref{sec_nearvoids}), in light red and blue
    colours, respectively.}
    \label{voiddist}
\end{figure}

Cells with $\rm \delta_{cell}<-0.8$ are defined as void centre candidates.
Over these candidates the algorithm centre spheres of variable radius, $r$, 
and computes the integrated density contrast profile as:
\begin{equation}
    \Delta(r) = \frac{n_{\rm gal}}{n_{\rm rand}}\,\frac{N_{\rm rand}}{N_{\rm gal}}-1\,,
\end{equation}
where $n_{\rm gal}$ and $n_{\rm rand}$ are the number of tracers galaxies and random points inside the sphere, $N_{\rm gal}$ and $N_{\rm rand}$ the total
number of galaxies and random points in the sample volume, respectively.
Void candidates are selected as those regions having $\Delta(r_\mathrm{void}) < -0.9$ 
at a given scale $r_\mathrm{void}$. The centres of the candidates are 
randomly perturbed around their initial position and recalculating its density profile $\Delta(r)$,
to look for the largest sphere having $\Delta(r_\mathrm{void}) < - 0.9$.
Finally, all overlapping spheres are removed keeping only the largest ones.
Fig. \ref{voiddist} shows the distribution of the radii of voids identified 
in the data (grey shaded area). As can be seen, the behaviour of the void abundance histogram
has a maximum around $10$\mph. Voids with sizes larger than this transition radius,
usually defined as the radius of the shot noise (vertical dashed black line),
behave as expected in $\Lambda$CDM cosmology: large voids are less abundant
than small ones. The opposite behaviour is seen to the left of the shot noise
radius. This lack of completeness is an expected effect of the shot noise in
the sample \citep[see for example][ and the references in]{Correa_vsf}.
Therefore, in this work we only consider voids with radii $\rm r_{void}>10 \, $\mph,
and at least $\rm 1.5\,r_{void}$ away from the boundaries of the catalogue.
Given these conditions, our final sample consists of 234 voids.
Following the classification of voids presented in \cite{Ceccarelli2013}, 
we separate voids into two categories: those in expansion at all scales, R-type,
and those expanding at their inner regions but collapsing at larger scales, S-type. 
This classification is performed based on the maximum value of $\Delta$ attained within the range of $1-3,r_{void}$. S-type voids correspond to values of $\Delta$ greater than zero, while R-type voids correspond to values of $\Delta$ less than zero.
The distribution of sizes of these two categories is shown in the Fig. \ref{voiddist}
using the light red and blue coloured shaded areas, as indicated in the figure key
(see more details in section \ref{sec_nearvoids}).

\subsection{Groups Identification}
\label{secciongruposcat}

We employ the galaxy group catalogue described in \cite{Rodriguez2022} compiled using the algorithm presented in \cite{Rodriguez2020}
on the SDSS DR16 data. This method is a combination of a Friend-of-Friends algorithm \citep[FOF;][]{Huchra1982} with a halo-based
method \citep{Yang2005}. As first step, the procedure to identify groups consists of linking galaxies with a FOF algorithm to find
gravitationally bound systems with at least one bright galaxy (i.e., a galaxy brighter than -19.5 in the r-band). 
Then, a three-dimensional density contrast in redshift space is calculated, with a characteristic luminosity estimated with the
potential galaxy members. To this calculation, the method considers the incompleteness produced by the limit magnitude of the
observational catalogue. 
Therefore, following \citet{Moore1993}, we add to the observed luminosity of the group a correction for the missing galaxy members.
Then, the halo mass of each group is assigned by abundance matching on luminosity \citep{Kravtsov2004, Tasitsiomi2004, Cristofari2019}
This procedure assumes a one-to-one relationship between characteristic luminosity and halo mass. It consist in sorting the halos by its luminosity and assigning masses according to the abundance prescribed by the mass function.
Using this mass and assuming that galaxies follow a distribution of an NFW profile, the algorithm computes the three-dimensional density contrast and
associates galaxies with groups. With this new membership assignment, the algorithm recalculates the characteristic luminosity and
iterates until it converges. This approach produces excellent results in terms of completeness and purity \citep{Rodriguez2020}.

We would like to emphasise that the masses of the galaxy groups determined by the algorithm give reliable results in comparison to those of weak gravitational lensing \citep{Gonzalez2021}.
On the other hand, the method has been successful in obtaining the Halo Occupation Distribution \citep[HOD; ][]{Rodriguez2020, Alfaro2022}, central galaxy alignment \citep{Rodriguez2022} as well as scaling relations \citep{Rodriguez2021}. 

\begin{figure}
	\includegraphics[width=\columnwidth]{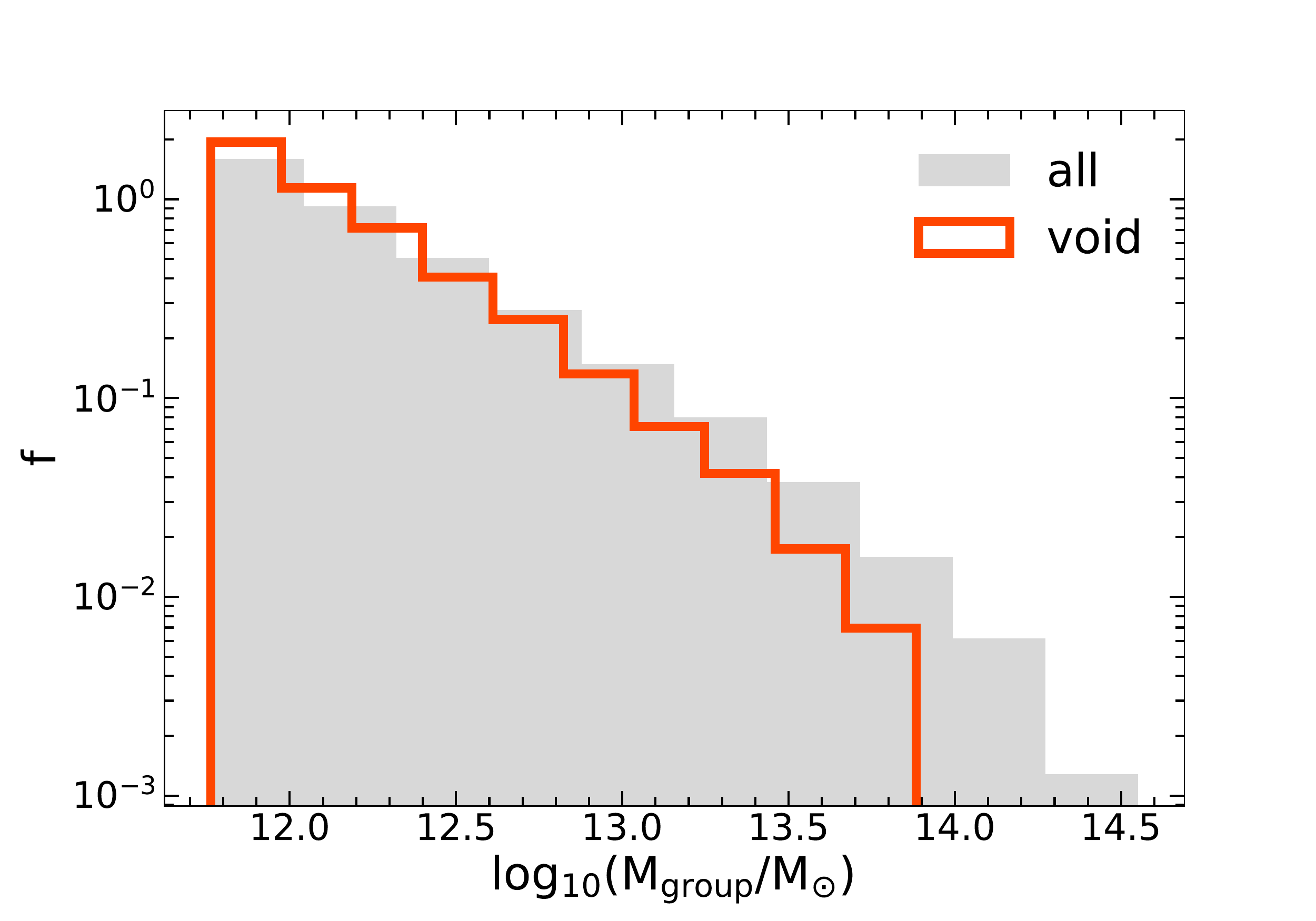}
    \caption{ Group mass distribution for haloes inside of cosmic voids (red-line ) and in the complete sample (grey-filled).
    The distribution show a lack of massive haloes in voids and that this is dominated by low-mass haloes.
    Our sample are composed by 109244 groups where 4094 are identified inside of cosmic voids.
    }
    \label{halodist}
\end{figure}

In Fig. \ref{halodist} we show the group mass distribution obtained for systems inside voids and in the general sample. 
As expected, we can see that the distribution corresponding to haloes in voids is dominated by low masses.
This is because the galaxy luminosity distribution in voids is composed of fainter galaxies.
Despite the lack of massive haloes in voids compared to those in the overall sample, we observe the presence of
some massive haloes with $\rm M_{group}\sim 10^{13.5}\, M_{\odot}$ inside voids.
When we consider the fraction of galaxies residing in groups, we find that the $\sim 73 \%$ of the galaxies are in groups, but 
for void galaxies, this percentage is $\sim 65 \%$. This show an excess of isolated galaxies in voids 
(compared with the general population). 
Nevertheless, we want to point out that the presence of galaxy groups is important in cosmic voids, even
if they are dominated by low-mass groups, as galaxy pairs and triplets.

\section{Void-galaxies properties}
\label{sec_voidgalaxies}

In this section, we study possible dependencies between the galaxy properties with the large-scale environment 
provided by voids by also controlling the local density around each galaxy.
To analyse the effects of the large-scale structure in galaxies, we separate them according to their 
belonging to a void. 
To do this, we calculate for each void the distance between its centre and all
nearby galaxies, defining as void galaxies those within the void radius $r_\mathrm{void}$.
As we mentioned in Subsection \ref{sec_void}, by definition there are no overlapping voids, 
and therefore each void galaxy belongs to only one void in the catalogue.
On the other hand, we define the general population of galaxies as the total sample of objects
regardless of whether or not they belong to a void.
It may be noticed that we compare void galaxies with the general sample (which also includes void galaxies) instead of with the sample of galaxies outside voids. As mentioned before, we are only taking into account those voids not affected by the survey boundaries (voids separated from the boundaries more than $1.5$ $r_\mathrm{void}$, see Section \ref{sec_void}). Thus by definition, the sample of galaxies in voids defined in the survey is not complete but pure. On the other hand, a sample of galaxies outside voids can not be defined in a pure way, therefore we restrict ourselves to the comparison of galaxies inside voids to the general population.

To describe the local density of a galaxy, we will use the group finder described in Section \ref{sec_data}.
Galaxies belonging to a group are by definition in a system with at least one bright galaxy, and the
algorithm employs the total luminosity of the group (using the galaxy members with $M_{r}<-19.5$) to
estimate the mass of the dark matter halo of the system. In this way, the halo
mass can be used as an estimation of the local density \citep[see for instance][]{Ceccarelli2012}. 
On the other hand, if a galaxy does not belong to any group is by definition a faint galaxy \textbf{(with $M_r>-19.5$)} in a region isolated from bright galaxies.
Therefore, although it is not possible to estimate the mass of the halo of these galaxies, they are expected
to be located in local environments of lower density.
As a first step in subsection \ref{sec:all}, we will analyse galaxies according to
their belonging to a void without considering their group membership.
Then in subsection \ref{sec:aisladas}, we focus the analysis to isolated galaxies, whereas in subsection
\ref{sec:grupos} we study the behaviour of the galaxy properties in voids and in the general population in bins of halo mass.

Regarding the galaxy properties mentioned above, we focus in particular, on the study of the $g-r$ colour
and the star formation rate (SFR).
These properties give us an indication of the effect of the void environment on the past
and present star formation processes in galaxies.
To study whether the void environment also influences the morphology of galaxies, we looked at the concentration parameter.
This is defined as $c=r_{90}/r_{50}$ where $r_{90}$ and $r_{50}$ are the radii containing the $90$ and $50$ per cent of the Petrosian flux, respectively. 
It is expected larger (lower) values of this parameter for early-type (late-type) galaxies.
Typically galaxies with $c>2.6$ are classified as elliptical galaxies \citep{Strateva2001}.
The galaxy properties studied in this work are strongly dependent on the stellar mass \cite[see, for e.g.,][]{Dickinson2003, Baldry2006,Peng2010}.
Since we are interested in the impact of the large-scale environment on the evolution of galaxies,
we compare their properties in the different samples as a function of stellar mass.

In this work we restrict our analysis of the different populations of galaxies to the mean values
(first moment) of their astrophysical properties. An analysis of higher order moments requires higher
levels of statistical significance and is beyond the scope of this work.

\subsection{General void-galaxies population}
\label{sec:all}

\begin{figure*}
	\includegraphics[width=\linewidth]{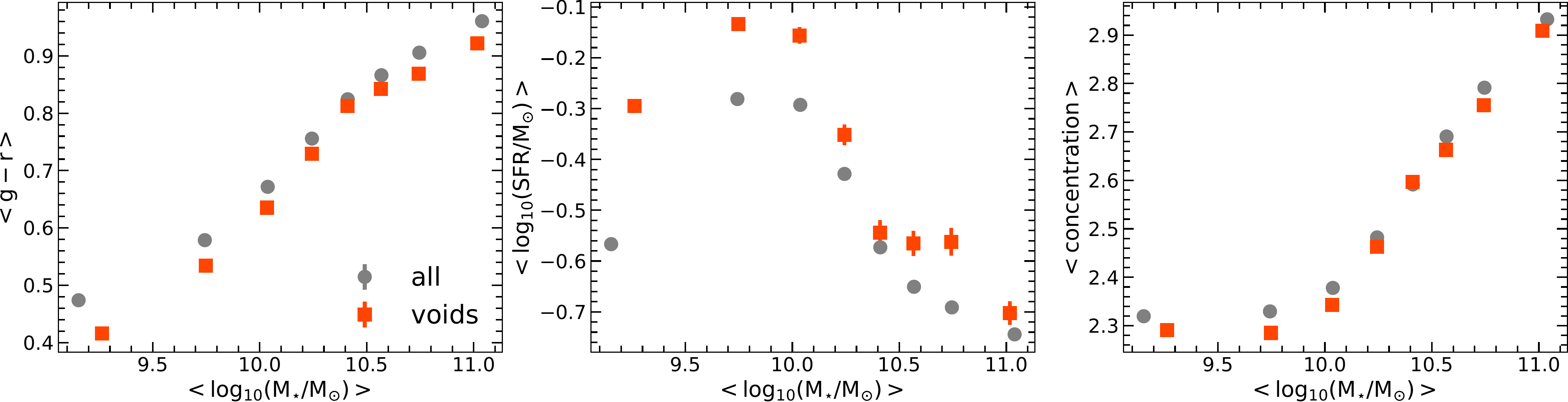}
    \caption{Mean values of galaxy properties as a function of the mean stellar mass of galaxies in stellar mass bins. The vertical bars represent the standard error. The orange squares correspond to galaxies in voids, while the grey circles represent the values of the general sample of galaxies.
    \textit{Left Panel:} mean $g-r$ colour. \textit{Central Panel:} mean of the logarithm of the star formation rate. \textit{Right Panel:} concentration parameter (i.e. $r_{90}/r_{50}$).
    In some cases, the error bars are not noticeable because they are smaller than the size of the marker.
    }
    \label{all}
\end{figure*}

To study the variation of the astrophysical properties of galaxies with the stellar mass,
we show in Fig. \ref{all} the mean $g-r$ colour (left panel), the star formation rate (central panel) and the concentration parameter of galaxies (right panel).
In grey circles, we show the complete sample of galaxies and, in orange squares, the sample of galaxies in voids. 
The vertical bars indicate the standard error of the mean for the galaxies in each bin of stellar mass.  
In some cases, the error bars are not observed because they are smaller than the marker.
For galaxies with $\sim M_{\star}<10^{10.3}M_{\odot}$ we found that those in void are bluer, with a higher
star formation rate and less concentrated than galaxies in the general sample. On the other hand, for
galaxies with higher stellar mass (i.e. $\sim M_{\star}>10^{10.3}M_{\odot}$) the trend in colour and
concentration seems to continue, however, the differences are within the error bar.
In the star formation rate, we found that galaxies in voids are more active than the general population
across the entire range of stellar masses.

These trends are consistent with those reported in previous work
\citep{Rojas2004,Patiri2006a,Ricciardelli2014}, where void galaxies were found to be more active in star formation compared to the general population of galaxies. Due to high concentration indices are related to elliptical morphologies, our results are also consistent with previous morphological studies of void galaxies, which found that the population of galaxies in the inner regions of voids are typically spiral galaxies \citep{Hoyle2005,Ricciardelli2017}.
These results can be interpreted as an effect of the void environment in galaxies. Nevertheless, these trends can also be attributed only to a local density effect, since the distribution of galaxy local densities in void-galaxies is expected to be biased towards lower densities \citep[see for instance][]{Gonzalez2009,RodriguezMedrano2022}.

\subsection{Isolated galaxies}
\label{sec:aisladas}

\begin{figure*}
	\includegraphics[width=\linewidth]{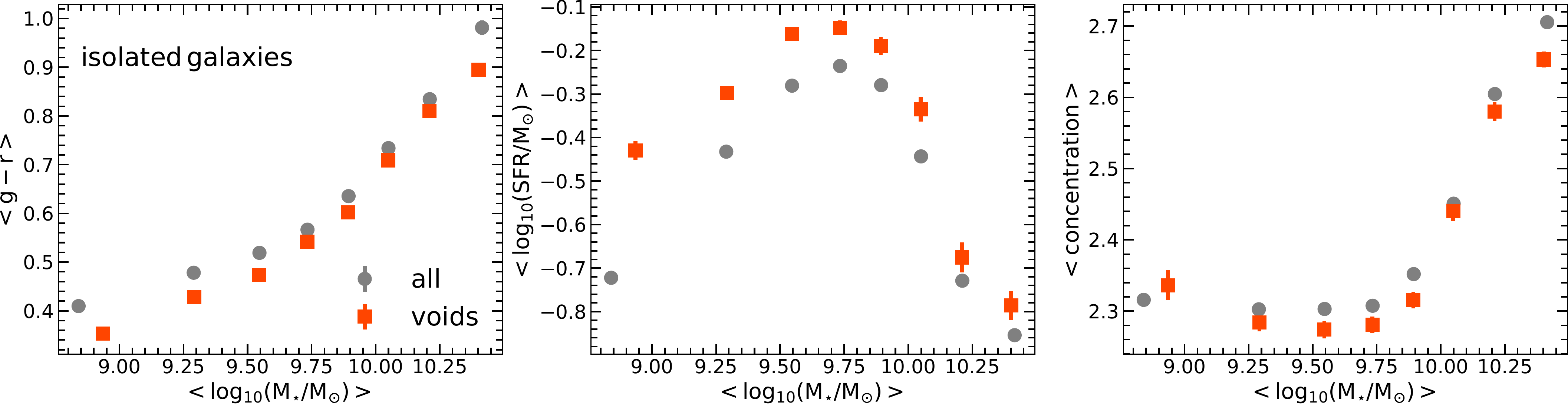}
    \caption{  
    On each panel it is shown the mean of different galaxies properties as a function of the stellar mass for void galaxies and the general sample for galaxies classified as \textit{Isolated}, this is galaxies not contained within a group. \textit{Left Panel:} g-r colour, \textit{Central Panel:} star formation rate, \textit{Right Panel:} concentration ($r_{90}/r_{50}$). In all the panels
   orange-squares are the values for void galaxies and grey-circles correspond to galaxies in the general sample. Error bars indicate the standard error.
   In some cases, the error bars are not noticeable because they are smaller than the size of the marker.
  }
    \label{aisladas}
\end{figure*}

As we mentioned at the beginning of this section, galaxies classified as isolated by the group finder are faint ($M_r>-19.5$) and do not have an estimate of the halo mass. However, for this study, we assume that isolated galaxies have a similar local density since they are not associated with any group. In Fig. \ref{aisladas} we show the properties of isolated galaxies according to the large scale environment. As in the previous subsection (see Fig. \ref{all}), we show the mean values of the $g-r$ colour index, the SFR and the concentration, from left to right. All these properties are shown as a function of stellar mass. The error bar corresponds to the standard error $\sigma$.
As can be seen, there is a trend indicating that galaxies within voids are bluer than galaxies in the overall sample. For stellar masses below $\sim 10^{9.8} M_{\star}$ the signal is stronger and the mean colour show statistically significant differences between galaxies in voids and the general population.
The medium panel shows that over all the stellar mass range, void galaxies are more star-forming than galaxies in the general sample, with stronger differences in some stellar mass bins. 
In the right panel of the figure, we found a tendency of the galaxies in voids to be less concentrated, and this difference is generally in the order of $\sim \sigma$.  

\subsection{Galaxies in groups}
\label{sec:grupos}

\begin{figure}
	\includegraphics[width=\columnwidth]{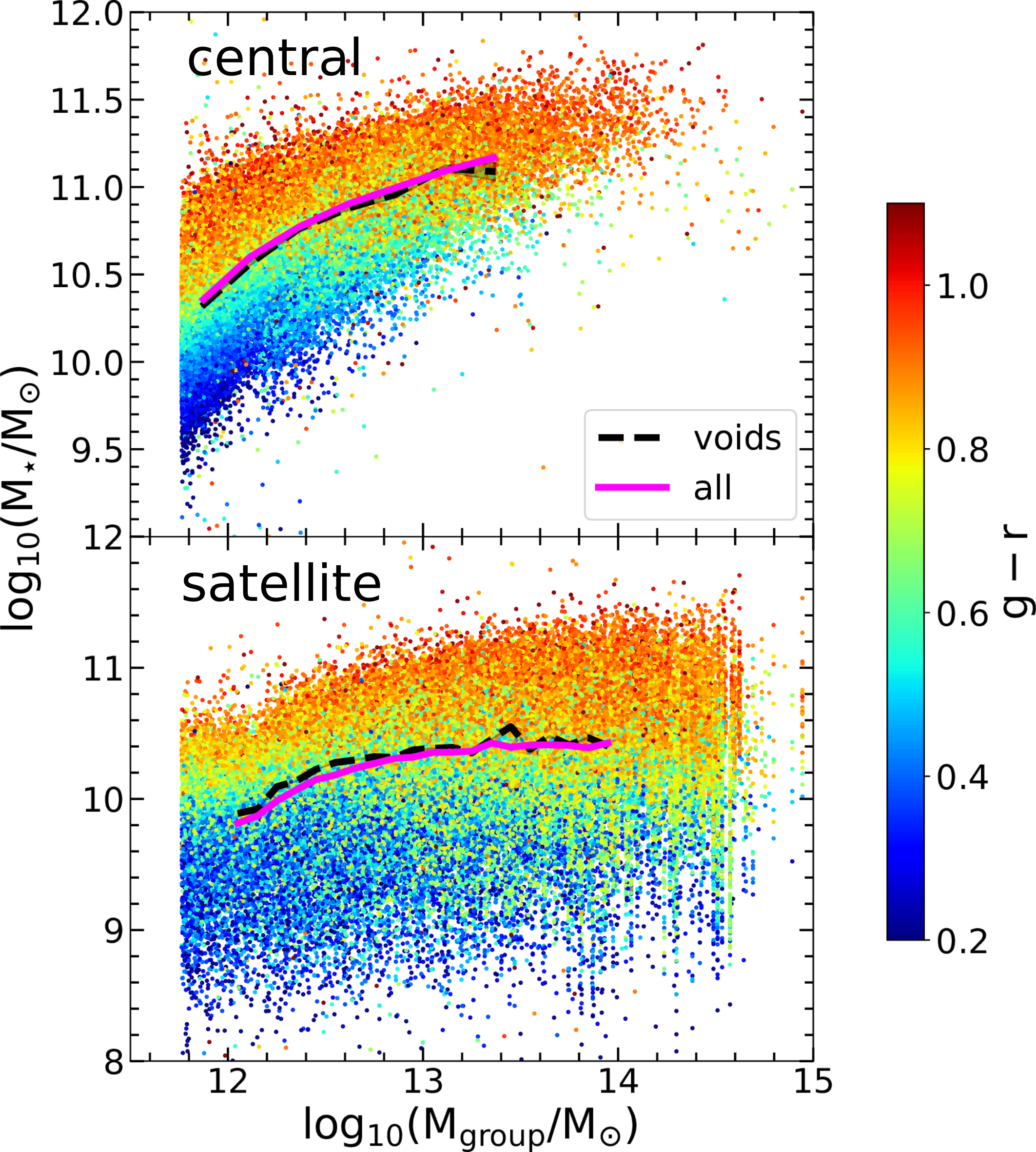}
    \caption{ Group mass versus stellar mass for central (top panel) and satellite (bottom panel) galaxies. Each point represent a galaxy and the colour map indicate the g-r galaxy colour. The solid magenta line indicate the mean stellar mass in the group mass bin and the dashed-black line indicate the mean stellar mass for void galaxies. The shaded area is the standard error.  }
    \label{abundancemat}
\end{figure}

In this section, we analyse galaxies classified as members of a group. We separate them into central and satellite galaxies.
The former is identified as the brightest galaxy of each group. 
To study how the stellar mass in galaxies is associated with the total mass of the group (hereafter halo mass), in Fig. \ref{abundancemat} we show a scatter plot of the stellar-to-halo mass relation for central (top-panel) and satellite (bottom-panel) galaxies,
regardless of whether or not they belong to a void. 
Each point in both panels represents a single galaxy coloured accordingly to its colour index $g-r$, as indicated in the figure key.
The black shaded line indicates the mean stellar mass for void galaxies and the magenta solid line is the mean stellar mass for all galaxies. 
In both cases, the estimate of the error in the mean is small, barely visible, but is shown as a shaded area.
In the case of the central galaxies, as can be seen, there are no statistically significant differences, however, there is a marginal tendency for the central galaxies of groups in voids to have a lower stellar content, at a fixed halo mass.
For satellite galaxies, we have the opposite trend. On average, for a given halo mass, satellite galaxies 
in voids are more massive. In this case, the differences between mean values are larger than the error estimates.

For central galaxies, in Fig. \ref{abundancemat} it can be seen that in a given stellar mass bin, galaxies in low-mass haloes are redder than galaxies in high-mass haloes. 
Assuming that colour is an indicator of galaxy age, 
this can be interpreted as a manifestation of the assembly bias \citep{Artale2018, Tojeiro2017, Zehavi2018}. 
Nevertheless, the colour index variation is stronger when we fix the halo mass and look at the index in galaxies of different stellar masses. This suggests that the colour of a galaxy depends mainly on the stellar mass, while the halo mass dependence is a second order relation \citep{Bamford2009}
Similarly, for satellite galaxies, the colour index seems to depend strongly on the stellar mass. In this case, the dependence on the halo mass is weak and appears to be present only for satellites with low stellar mass in massive haloes.

\begin{figure*}
	\includegraphics[width=1.5\columnwidth]{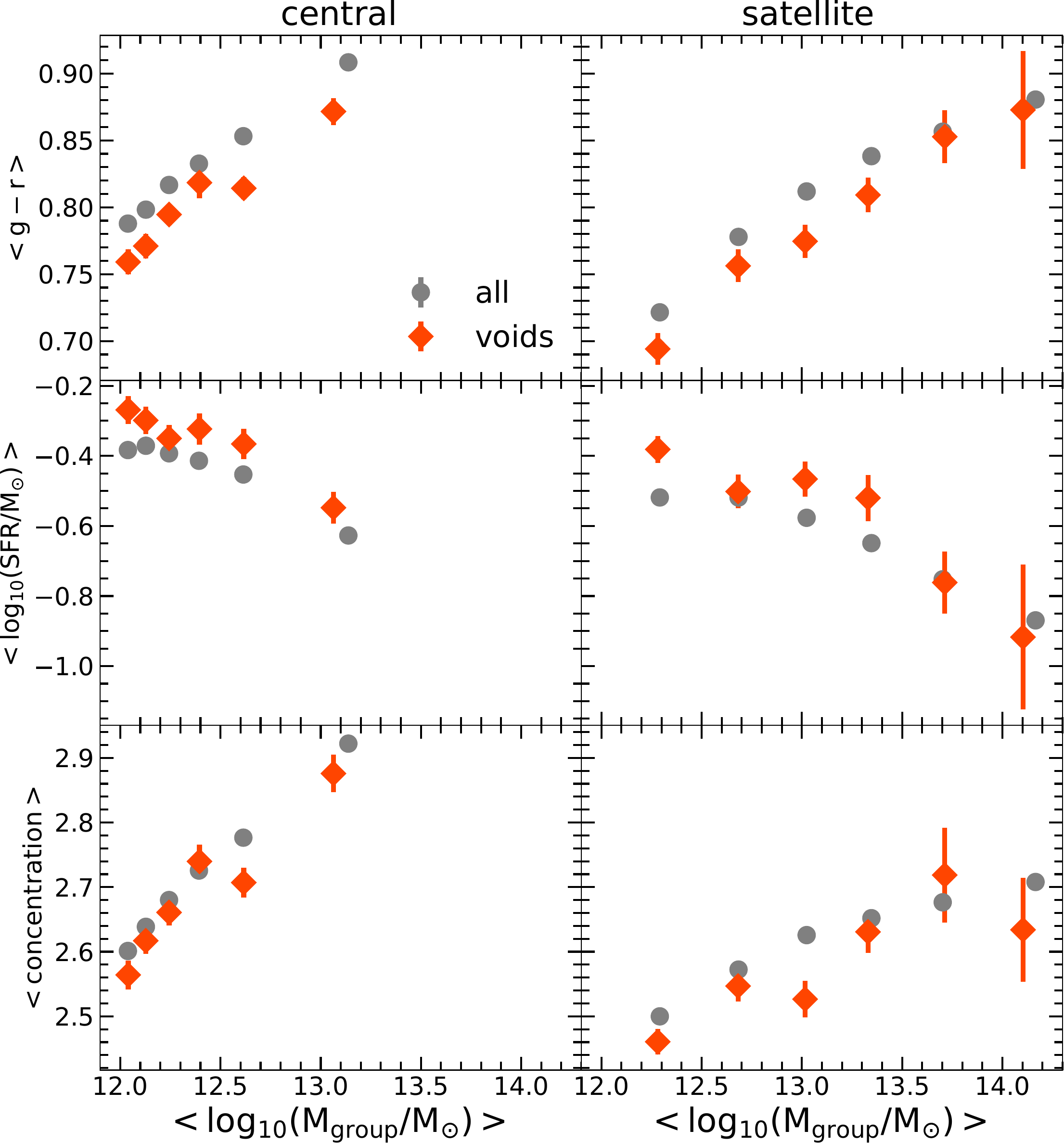}
    \caption{ 
    Mean value of different galaxy properties for central (left panels) and satellite galaxies (right panels). 
    In the top panels the $g-r$ index, in the central panels the SFR and in the bottom panels the concentration parameter. With orange-diamonds we show the void galaxies and in grey-circles all the galaxies in the sample. The vertical bars represent the standard error.
    In some cases, the error bars are not noticeable because they are smaller than the size of the marker.
    }
    \label{centrales}
\end{figure*}

To analyse the relation between galaxies in groups inside cosmic voids and in the general universe, 
we construct different halo mass bins with equal number of galaxies for each sample.
As we mentioned before, halo mass is used as a proxy of the local environment, in order to control possible dependencies with the large scale environment provided by voids.
In Fig. \ref{centrales} we show the results for central (left panels) and satellite (right panels) galaxies belonging to haloes in voids (orange diamonds) and in the general population (grey circles, as indicated in the figure key). For each sub-sample we
show the mean halo mass in the abscissas while in the ordinates of each panel we show the mean values of different galaxy properties, that is g-r, star formation rate and concentration from top to bottom. 
For satellite galaxies, these mean values are weighted by the inverse of the number of galaxy members in the halo. 
The error bars are given by the standard deviation ($\sigma$).
In both environments, inside and outside of voids, we observe the expected trend for galaxy properties as a function of halo mass, that is the mean concentration and colour increase as halo mass increases, whereas as occurs the opposite with the mean star formation rate \citep{Postman1984,Kauffman2004,Balogh2004b,Blanton2005}.
Also it can be seen that central galaxies are redder, more star-forming and more concentrated than satellite galaxies.
More relevant in the context of this work are the dependencies of the signal on the large-scale environment, independent of the mass of the halo.
For haloes with masses lower than $\sim 10^{13.5}$ \msol, the g-r colour index and star formation rate signals are highly dependent on the environment. Central and satellite galaxies in voids are bluer than the general population. They are also more star-forming than galaxies in the general universe.
For the halo mass - concentration relation, galaxies in voids are less concentrated than their counterparts, especially for satellite galaxies and halo masses lower than $\sim 10^{13.5}$ \msol.
Note that, as a consequence of dividing the sample into halo-mass bins with the same number of galaxies, the mass ranges covered for the central and satellite galaxies are different.
This is because all haloes have central galaxies by definition, whereas satellite identification is more likely for massive haloes.
Thus, the sample of central galaxies is dominated by low-mass haloes whereas for satellites a better sampling of high-mass haloes is obtained.

We have analysed above the existence of correlations of galaxy properties with the large scale environment provided by voids while the local environment is controlled by means of galaxy groups.
As we show in Fig. \ref{abundancemat}, there is no clear dependence of the stellar to halo mass relation on the environment for central galaxies, and in the case of satellite galaxies such dependence, although significant, is very weak. However, it remains to verify whether the dependencies
of galaxy properties within the large scale environment could be driven by the galaxy stellar mass.

\begin{figure}
	\includegraphics[width=\columnwidth]{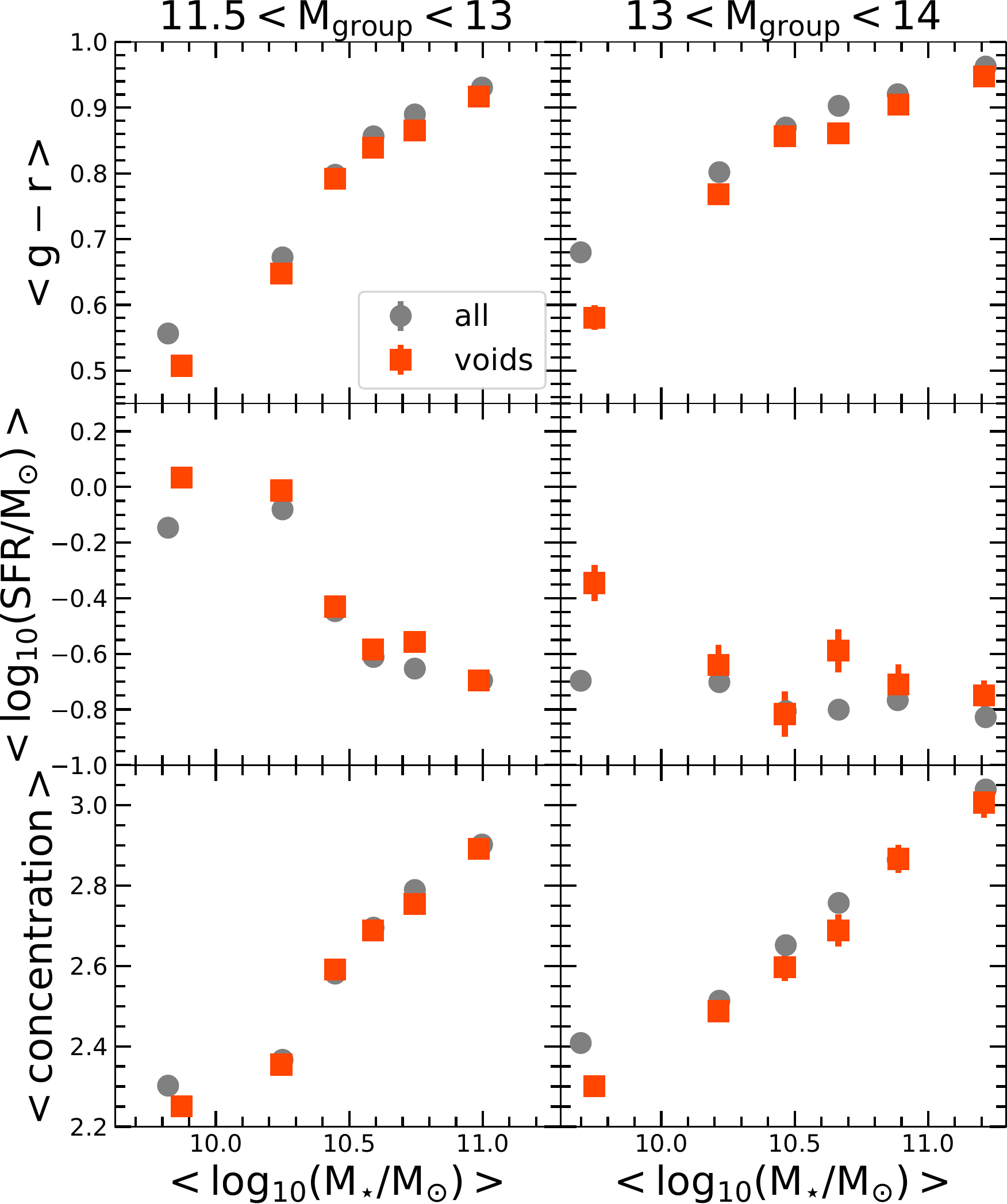}
    \caption{ Mean colour, SFR and concentration for galaxies in groups, as a function of the stellar mass. In left panels we show the relations for galaxies in groups with  $\rm 11.5<M_{halo}<10^{13}\,M_{\odot}$. In the right panels for galaxies with $\rm 10^{13}<M_{halo}<10^{14}\,M_{\odot}$ In orange-squares we show void galaxies and in grey-circles the general sample of galaxies.
    In some cases, the error bars are not noticeable because they are smaller than the size of the marker.
    }
    \label{gruposxmst}
\end{figure}

In Fig. \ref{gruposxmst} we show the mean values of colour, SFR and concentration parameter for galaxies samples in groups as a function of their stellar mass. 
Here the approach is similar to the one used in Fig. \ref{all} and Fig. \ref{aisladas} and complementary with the one used in Fig. \ref{centrales}. 
In the left and right panels of Fig. \ref{gruposxmst} we show the mean properties of galaxies in groups with masses in the intervals $\rm 10^{11.5}<M_{halo}<10^{13}\,M_{\odot}$ and $\rm 10^{13}<M_{halo}<10^{14}\,M_{\odot}$, respectively. 
The overall comparison between the two group mass ranges shows that the more massive groups (right panel) generally have redder, lower star forming, and more concentrated galaxies.
The comparison between galaxies in voids or in the overall sample, regardless of halo mass, shows the same trends as those observed in Fig. \ref{centrales}.
Void galaxies are bluer, with higher star formation rates and lower concentration than in the overall sample.
This trend is present throughout the entire range of stellar masses analysed.
In this analysis for brevity we do not split the sample into central and satellite galaxies, as we find that the signal is consistent in both types.

\section{Galaxies in the near void regions}
\label{sec_nearvoids}

As was introduced in Sect. \ref{sec_data}, the integrated density contrast profile ($\Delta(r)$) allow us classifiy voids into two categories: those embedded in overdense surrounding regions and those in a underdense regions. The former are called S-type voids (\textit{shell profile}) while the latter are R-type voids (\textit{rising profile}).
This dichotomy in the integrated density contrast profile in voids was first detected in the observations by \citet{Ceccarelli2013} and predicted in simulations by \citet{Sheth2004}.

\citet{RodriguezMedrano2022} found differences in the haloes population inside and outside between R-S type voids.
Haloes in R-type voids grow more slowly than haloes in S-type voids in numerical simulations.
This different evolutionary paths could produce differences in the properties of galaxies, such as colour and star formation.  
To study this effect in our observational dataset, we split the voids sample into R-S type according to the highest $\rm \Delta_{max}$ reached between $ \rm 2-3\,r_{void}$ \citep{Ceccarelli2013}.
We consider only those voids that are at least $ \rm 2\,r_{void}$ from the edge of the catalogue.
As the R-S type classification is environment sensitive, we changed this criterion compared to the previous section, where we were interested in galaxies within voids.
Therefore, our current sample consist of 102 S-type and 66 R-type. 
In Fig. \ref{voiddist} we show in blue and red the R-S type void radii distributions respectively.
A comparison with the overall void sample (grey histogram) shows that R-type voids contribute more to the large-size sample, while S-type voids populate the small-size sample with a peak at $\sim 10 h^{-1}\rm Mpc$.

We constructed profiles of mean colour, SFRs (specific star formation rate) and concentration to study the dependence on the radial distance influence the galactic properties as a function of the void type.
In this section, we use the SFRs to avoid differences in the mass distributions in each radial that is noun correlates with the SFR \citep{Dave2008}.
Due to the void identification algorithm, it is expected to find the brightest galaxies outside the void. Then, to reduce the bias introduced by the lack of bright galaxies in the inner regions of the voids, we selected only galaxies fainter than $\rm M_{r}=-19.77$ and in a sample complete in volume. 
The analysis in this section was performed on three samples complete in volume up to $z<0.5\,(\rm M_r<-18.1)$, $z<0.65\,(\rm M_r<-18.6)$ and $z<0.8\,(\rm M_r<-19.1)$. We found consistent results in all cases, therefore, for simplicity, we only show the results corresponding to the sample up to $z=0.65$.

\begin{figure}
	\includegraphics[width=\columnwidth]{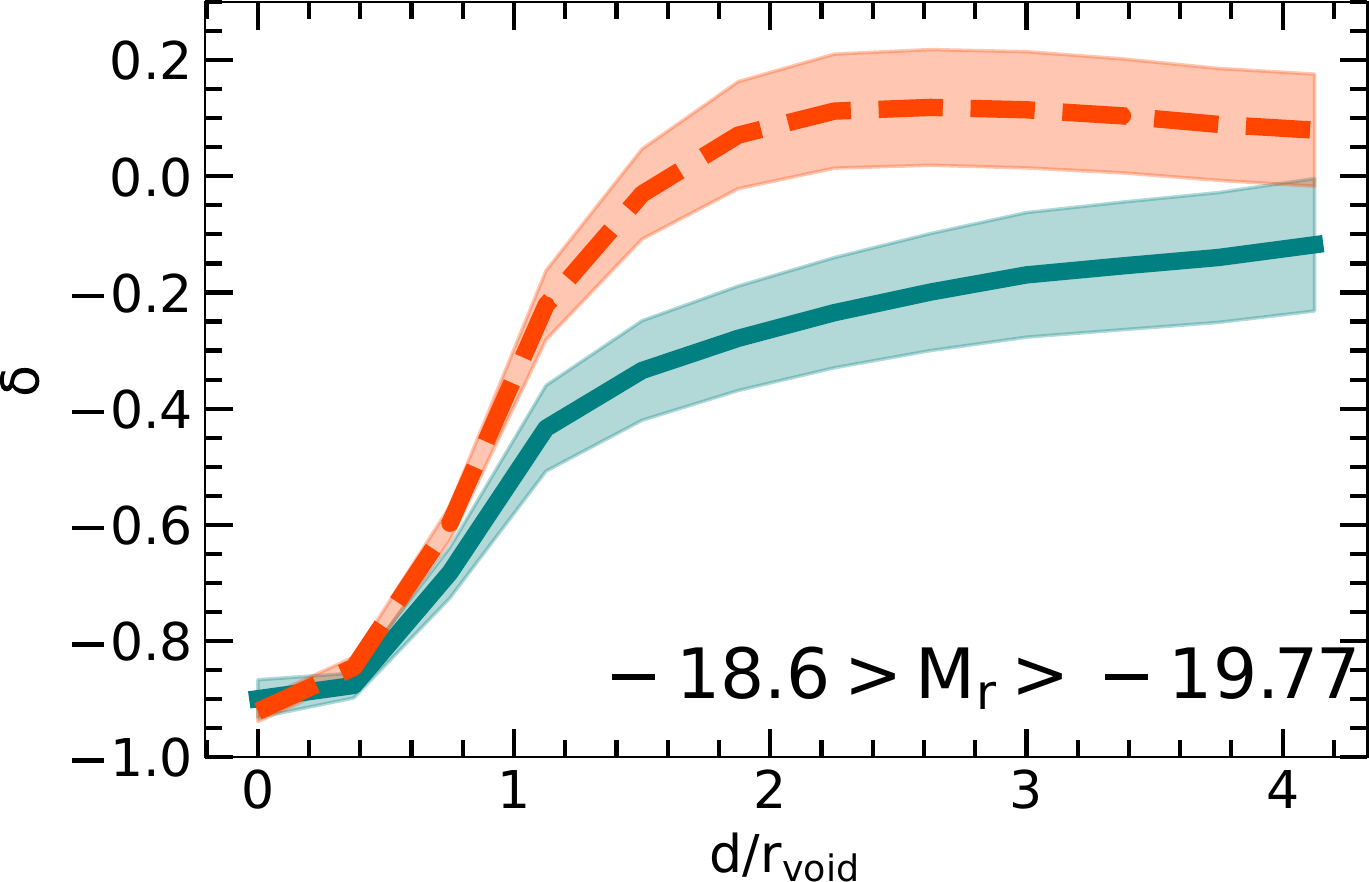}
    \caption{ Contrast density profile of galaxies. The profile were constructed on a sample with galaxies in the magnitude range of $-18.6> \rm M_r > -19.77$. In dashed red line the S-type void profile and in solid green the R-type. The shaded area is the standard error. 
    }
    \label{perfiles_delta}
\end{figure}

First, to show the densities of galaxies in the function of the void centre, we calculated the contrast density profile, defined as
\begin{equation}
    \delta(r) = \frac{\rho(r) - \hat{\rho}}{\hat{\rho}}
\end{equation}
where $\rho(r)$ is calculated as the number of galaxies in a shell at a distance $r$ from the centre of the void divided by the volume of the shell. Similarly, the $\hat{\rho}$ is calculated with a mask of random points.
Fig. \ref{perfiles_delta} shows the contrast density profile ($\delta$). Beyond $\rm r_{void}$, S-type voids have a $\delta>0$ indicating that these voids are over-dense and the opposite occurs for R-type voids, indicating that these are under-dense.

\begin{figure}
	\includegraphics[width=\columnwidth]{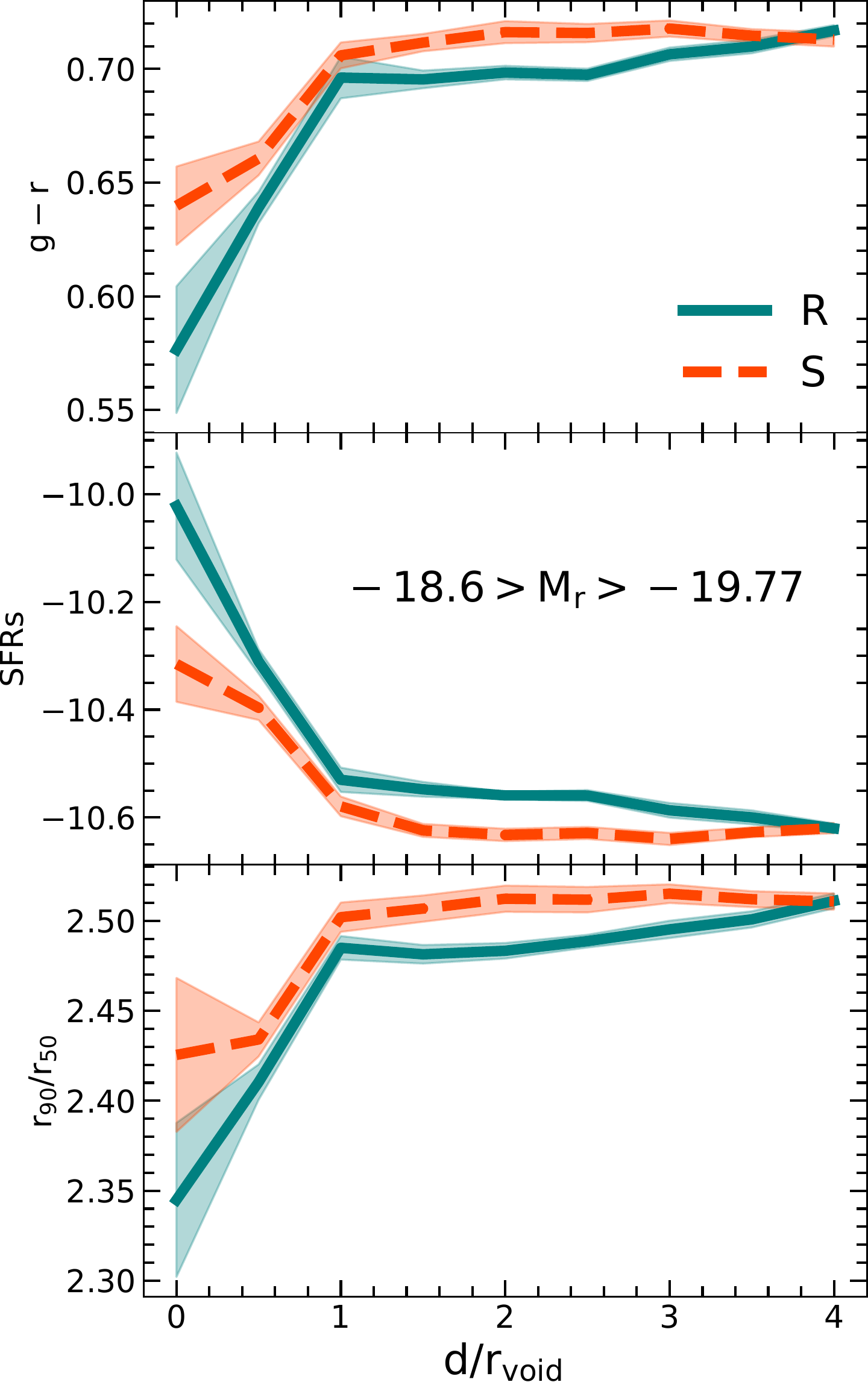}
    \caption{ Properties of galaxies as a function of the distance to the void radius. \textit{Top Panel:} mean g-r colour \textit{Medium Panel:} sSFR, \textit{Bottom Panel:} concentration $r_{90}/r_{50}$. In dashed-red colour we show the S-type void profile and in solid green line the R-type void profile. The profiles were constructed on a sample with galaxies in the magnitude range of $-18.6>\rm M_r >-19.77$.}
    \label{perfiles_prop}
\end{figure}

Fig. \ref{perfiles_prop} shows the mean colour, SFRs and concentration profiles in the same sample, in red-dashed (solid blue) are S-type (R-type). The shaded area indicates the estimate of the standard error ($\sigma$).
Regardless of the void type, the inner regions are bluer, with a higher star formation rate and less concentrated galaxies than the outer regions. 
When comparing between void types, we found that R-type voids have bluer, more star-forming, and less concentrated galaxies than S-type void in the inner regions.
Due to that typically young galaxies have blue colours, high star formation rates and low concentrated morphologies \citep[e.g.,][]{Strateva2001,Brinchmann2004,Baldry2006}, the results of the figure support the idea of a different evolution path with the void type \citep{RodriguezMedrano2022}.

Beyond $\rm r_{void}$, S-type galaxies are redder than R-type galaxies. At a distance close to $\sim 4\, \rm r_{void}$ the quantities studied reach the same value, indicating the distance of influence of the void environment.

\section{Discussion}
\label{sec_discusion}

In this work, we analysed galaxies in voids and in the general universe by controlling the local density.
When we consider isolated galaxies in voids, we found significantly bluer colours when the stellar mass is lower than $\sim 10^{9.75} \rm M_{\odot}$.
To study the local density dependence we select galaxies with equal host halo mass. We separate them into central and satellite galaxies.
For galaxies that inhabit groups, all the central galaxies in void-groups are bluer than central galaxies in denser large scale environments. For groups with masses below $\sim 10^{13.5}M_{\odot}$ satellite populations are also bluer than the total population. 
In higher mass groups, we have found no signs of environment dependence for satellite galaxies. 

The colour differences mentioned above are consistent with the behaviour of the SFR. Galaxies in voids are more star-forming, in comparison with other large scale environments, when equal local density is considered. In this case the trend is statistically significant in the whole mass range. 
For satellite galaxies, we found no significant differences for higher mass haloes (masses higher than $\sim 10^{13.5}$). 
The results clearly show the degree of influence of the large-scale void environment and the local density,
in agreement with previous studies \citep[see for instance ]{Rojas2004, Patiri2006b, vonBenda2008, Ricciardelli2014, Ricciardelli2017}. However, in those works the local environment is not controlled.

Our results in the colour, SFR and concentration, can be interpreted as if void haloes assemble later delaying the galaxies formation process. 
This interpretation is consistent with the previous works made with numerical simulation \citep{Hahn2007a,Hahn2007b,Tonnensen2015, Tojeiro2017,Martizzi2020,Alfaro2020} and observational data \citep{Douglass2018}.
In this way, void-galaxies are at the present time bluer, with higher SFR and have a less concentrated morphology than galaxies with equal local density in the general universe.

In addition to the astrophysical properties of the galaxies, we calculate the mean stellar mass as a function of the halo mass. For central galaxies, we found a tendency for void galaxies to inhabit larger dark matter haloes. This is consistent 
with previous studies on numerical simulations \citet{Alfaro2020,Rosas-Guevara2022}. 
Is important to note that these works use similar void definitions to the used here.

To deeply understand how the void environment affects galaxies, we analyse the profile of galaxy properties accordingly to the distance to the void centre. This profiles are computed up to a distance equivalent to four void radius. In this we define two regions for each void, galaxies inside the voids and galaxies in what we call the void outside, between one and four void radius.
We classify voids according to the behaviour of its integrated contrast density profile, separating in S and R type voids. 
Regardless of the void region, galaxies in S-type voids are redder, passive, and more concentrated than
galaxies in R-type voids. We interpret the quench signals found for galaxies in the outer region of S-type
voids as a consequence of higher density. In other words, the outskirts of S-type voids are dominated by
high-mass haloes that quench galaxies efficiently.

However, it is not clear what mechanisms act in the inner regions of the voids, depending on the type of
void, affecting the properties of the galaxies.
More detailed studies are necessary to fully understand the evolution of galaxies in void environments. 

Previous works have shown that the surroundings of voids affect star formation and morphology of galaxies depending on the void size.
Moreover, larger voids have an excess of late-type star-forming galaxies in their walls compared to the general universe, and the opposite is true for the smaller, as shown in \citet{Ricciardelli2014, Ricciardelli2017}. 
These authors suggest that the differences between the galaxy properties in void walls can be a manifestation of the dichotomy between S and R-type voids since the former are typically smaller \citep{Ceccarelli2013}. 
As mention before, the S-type voids have overdense surroundings where systems of galaxies and FVS (Future-Virialized-Structures) \citep{Lares2017} are located, explaining the excess of redder galaxies.

Our results on the astrophysical properties of galaxies as a function of the distance to void centre, are complementary to the analysis presented in \citet{Ricciardelli2014,Ricciardelli2017}. 
The separation of voids into small and large size is qualitatively consistent with the separation into S-type and R-type, where S-types show similar results to small voids and R-types to large voids.
This supports the idea, suggested by these authors, that the differences in the colour and morphology in the void walls are related with their environment.

\section{Summary and Conclusion}
\label{sec_conclusions}

We analysed a sample of galaxies belonging to the inner regions of spherical voids identified in the SDSS DR16. 
We compare the g-r colour index, the star formation rate and the concentration parameter of these void galaxies with the general population. 
We found that void galaxies are more star-forming, bluer and less concentrated than galaxies in the general universe.
To differentiate between the effects of the large-scale environment provided by voids and the local density, we identify galaxy groups and use their halo mass as an estimate of local density. 
For isolated galaxies with masses $\sim M_{\star}<10^{9.8}M_{\odot}$, the mentioned signals in galaxy properties persist. In the case of galaxies in groups, the mean stellar mass as a function of the halo mass varies with the large scale environment. For central galaxies, we find a tendency for void galaxies to inhabit larger dark matter haloes at a given stellar-mass bin. For satellite galaxies, we find the opposite trend. 

To determine if the quenching process affects in the same way the different classes of galaxies we 
studied the colour, SFR and concentration as a function of the halo mass. For central galaxies, we find
strong differences in colour and SFR between galaxies in void and in the general population. 
These trends are not statistically significant in the case of the concentration. For satellite galaxies, the signals are consistent but statistically marginal. 

To extend the analysis of the galaxies in the surrounding void regions, we studied the mean colour, SFRs and concentration, as a function to the void centre distance.
We find that the properties of galaxies depend on the void type. S-type voids have beyond the $\rm r_{void}$ a population of galaxies redder, less active and more concentrated than R-type voids. 
The void type also affects the galaxies in the inner region ($\rm d<r_{void}$) with the same tendencies.

To conclude, our results relate the large-scale void environment with the galaxy properties within them.
The existence of correlations on these properties with the void regions, while the local density is constrained, suggests a possible influence of the large-scale structure on the galaxy evolution.
Voids are expanding regions, where the formation of structures in their interiors is suppressed.
In this way, our results can be interpreted as an indication of a late evolutionary path of galaxies in voids
in comparison with the general Universe. On the other hand, the observed variation of galaxy properties in the inner part of voids in function of the void type, suggests a connection between the large-scale void dynamics and the galaxy evolution.

\section*{Acknowledgements}
This work was partially supported by the Consejo de Investigaciones Científicas y Técnicas de la República Argentina (CONICET) and the Secretaría de Ciencia y Técnica de la Universidad Nacional de Córdoba (SeCyT).
ARM is doctoral fellow of CONICET. DJP, FAS and ANR are members of the Carrera del Investigador Científico (CONICET).
The authors thanks support by grants PIP 11220130100365CO, PICT-2016-4174, PICT-2016-1975, PICT-2021-GRF-00719 and Consolidar-2018-2020, from CONICET, FONCyT (Argentina) and SECyT-UNC.
Funding for the Sloan Digital Sky 
Survey IV has been provided by the 
Alfred P. Sloan Foundation, the U.S. 
Department of Energy Office of 
Science, and the Participating 
Institutions. 

SDSS-IV acknowledges support and 
resources from the Centre for High 
Performance Computing  at the 
University of Utah. The SDSS 
website is www.sdss.org.

SDSS-IV is managed by the 
Astrophysical Research Consortium 
for the Participating Institutions 
of the SDSS Collaboration including 
the Brazilian Participation Group, 
the Carnegie Institution for Science, 
Carnegie Mellon University, Center for 
Astrophysics | Harvard \& 
Smithsonian, the Chilean Participation 
Group, the French Participation Group, 
Instituto de Astrof\'isica de 
Canarias, The Johns Hopkins 
University, Kavli Institute for the 
Physics and Mathematics of the 
Universe (IPMU) / University of 
Tokyo, the Korean Participation Group, 
Lawrence Berkeley National Laboratory, 
Leibniz Institut f\"ur Astrophysik 
Potsdam (AIP),  Max-Planck-Institut 
f\"ur Astronomie (MPIA Heidelberg), 
Max-Planck-Institut f\"ur 
Astrophysik (MPA Garching), 
Max-Planck-Institut f\"ur 
Extraterrestrische Physik (MPE), 
National Astronomical Observatories of 
China, New Mexico State University, 
New York University, University of 
Notre Dame, Observat\'ario 
Nacional / MCTI, The Ohio State 
University, Pennsylvania State 
University, Shanghai 
Astronomical Observatory, United 
Kingdom Participation Group, 
Universidad Nacional Aut\'onoma 
de M\'exico, University of Arizona, 
University of Colorado Boulder, 
University of Oxford, University of 
Portsmouth, University of Utah, 
University of Virginia, University 
of Washington, University of 
Wisconsin, Vanderbilt University, 
and Yale University.

\section*{Data Availability}

The data underlying this article will be shared on reasonable request to the corresponding authors.



\bibliographystyle{mnras}
\bibliography{paper} 





\bsp	
\label{lastpage}
\end{document}